\documentclass[letterpaper]{article}
\usepackage{aaai19}  
\usepackage{times}  
\usepackage{helvet}  
\usepackage{courier}  
\usepackage[hyphens]{url} 
\usepackage{graphicx}  
\usepackage{placeins}

\usepackage{colortbl}
\usepackage{nth}
\usepackage{capt-of}
\usepackage{comment}
\usepackage{multirow}
\usepackage{floatrow}
\usepackage{float}
\usepackage{booktabs}
\usepackage{enumitem}
\usepackage{multicol}
\usepackage{graphicx}
\usepackage{subcaption}
\usepackage{amsmath}

\usepackage{arydshln}

\newcommand\blfootnote[1]{%
  \begingroup
  \renewcommand\thefootnote{}\footnote{#1}%
  \addtocounter{footnote}{-1}%
  \endgroup
}

\begin{document}
\title{NELA-GT-2020: A Large Multi-Labelled News Dataset for The Study
of Misinformation in News Articles}

\author{Maur\'{i}cio Gruppi\textsuperscript{*}, Benjamin D. Horne\textsuperscript{\dag} and Sibel Adal{\i}\textsuperscript{*}\\
 Rensselaer Polytechnic Institute\textsuperscript{*}, University of Tennessee Knoxville\textsuperscript{\dag}\\
 gouvem@rpi.edu, bhorne6@utk.edu, adalis@rpi.edu
}

\maketitle

\begin{abstract} 
In this paper, we present an updated version of the \texttt{NELA-GT-2019} dataset~\cite{gruppi2020nela}, entitled \texttt{NELA-GT-2020}. \texttt{NELA-GT-2020} contains nearly 1.8M news articles from 519 sources collected between January 1st, 2020 and December 31st, 2020. Just as with \texttt{NELA-GT-2018}~\cite{norregaard2019nela} and \texttt{NELA-GT-2019}, these sources come from a wide range of mainstream news sources and alternative news sources. Included in the dataset are source-level ground truth labels from Media Bias/Fact Check (MBFC) covering multiple dimensions of veracity. Additionally, new in the 2020 dataset are the Tweets embedded in the collected news articles, adding an extra layer of information to the data. The \texttt{NELA-GT-2020} dataset can be found at: \url{https://doi.org/10.7910/DVN/CHMUYZ}.
\end{abstract}

\section{Introduction}\blfootnote{Data at: \url{https://doi.org/10.7910/DVN/CHMUYZ}} 
The study of news production behavior of both high veracity and low veracity outlets continues to be a salient research topic. One of the continued challenges in this interdisciplinary research area is having broadly labeled datasets over extended periods of time. For example, mixed-methods work on a variety of news-related research questions require data surrounding specific events~\cite{an2009news,horne2020tailoring}, which are often difficult to get post-event. Another example is machine learning research, which requires large, diverse, labeled datasets to train and test models~\cite{baly2018predicting,baly2020we,horne2019robust}. 

There has been numerous efforts to collect news article data to fill this need. These include the previous versions of the NELA dataset~\cite{gruppi2020nela,norregaard2019nela,horne2018sampling}, the FA-KES dataset \cite{salem2019fa}, and the Golbeck et al. dataset  \cite{golbeck2018fake}. The vast majority of datasets focused on information veracity have been social media posts rather than news articles, including datasets such as the FakeNewsNet dataset \cite{shu2018fakenewsnet}, and the LIAR dataset \cite{wang2017liar}. 

More recently, there has been a significant effort in creating COVID-19 information datasets, both for news articles and social media posts. Some of these datasets have veracity labels, others do not. For example, \cite{shahi2020fakecovid} created a news article dataset with fact-check labels, \cite{patwa2020fighting} manually annotated a set of social media posts and articles, and \cite{memon2020characterizing} created an annotated set of Tweets focused on COVID-19. 

 In this paper, we present the latest version of the NELA-GT datasets, \texttt{NELA-GT-2020}. \texttt{NELA-GT-2020} fills both needs described above, as it contains labeled news sources that covered a broad set of events, including the COVID-19 pandemic and the 2020 U.S. Presidential Election. More precisely, we have collected \textbf{1,779,127 news articles} from \textbf{519 sources} between \textbf{January 1st, 2020} and \textbf{December 31st, 2020}. Additionally, we include source-level veracity labels from Media Bias/Fact Check (MBFC) and embedded Tweets found in the news articles. 
 
 In this short paper, we describe the differences between the previous versions of the NELA-GT datasets and \texttt{NELA-GT-2020}. We also describe in detail the data collection method, ground truth collection method, embedded tweet collection, and publicly available data formats. Lastly, we provide some metadata and a discussion of potential use cases. 

\section{What's New in NELA-GT-2020?}~\label{sec:whatsnew} 
In addition to the updated time frame, there are several new additions to \texttt{NELA-GT-2020}:
\begin{enumerate}
    \item \textbf{More data}: The \texttt{NELA-GT-2020} dataset contains 1,779,127 news articles from 519 sources published in the year 2020, up from 261 sources in \texttt{NELA-GT-2019}. The sources added to the collection are mostly fringe, unreliable news outlets. Moreover, past versions of the \texttt{NELA} dataset focused on political news content. This time we have expanded our news collection to a broader range of topics, by collecting health-related news articles as well as general news feeds provided by the sources.
    \item \textbf{Embedded Tweets}: A new piece of information available in \texttt{NELA-GT-2020} are the embedded tweets. These consist of tweets featured in news articles. We provide the tweet content, author, date, and URL. We collected 410,432 embedded tweets across all of the news articles.
    \item \textbf{Ground Truth change}: In the 2020 dataset, we aggregated the ground truth information as provided by Media Bias Fact Check (MBFC). Concretely, we assigned reliability labels to news sources based on MBFC's Factuality Score. The labels are one of \emph{unreliable}, \emph{mixed}, or \emph{reliable}. 
\end{enumerate}

\section{Data Collection}
The data collection process follows what was described in \cite{norregaard2019nela}. 
Specifically, we scraped the RSS feeds of each source in our source collection list twice a day starting on 01/01/2020 using the Python libraries feedparser and goose. 
Our list of sources to collect was carried over from \cite{norregaard2019nela}, with an additional 258 sources added to this list. These additional sources are mostly conspiracy-driven and pseudoscience media sites from the Media Bias Fact Check list \footnote{\url{https://mediabiasfactcheck.com/}}. 
Just as in the 2018 and 2019 versions, these sources come from a variety of countries (or the country of origin is not known), but are all articles are in English.

\subsection{Data outage}

\begin{figure*}
    \centering
    \begin{subfigure}{0.6\textwidth}
        \includegraphics[width=\textwidth]{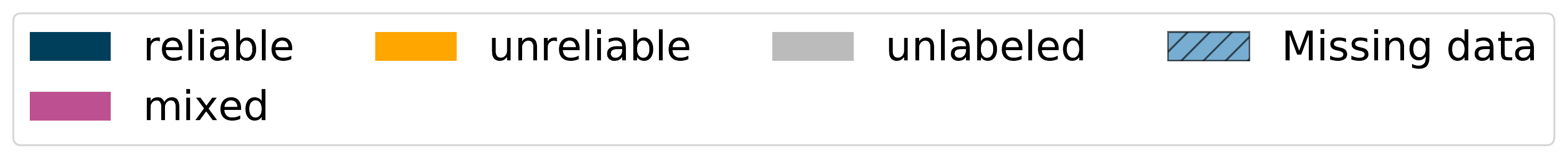}
    \end{subfigure} ~
    \begin{subfigure}{0.48\textwidth}
    \includegraphics[width=\textwidth]{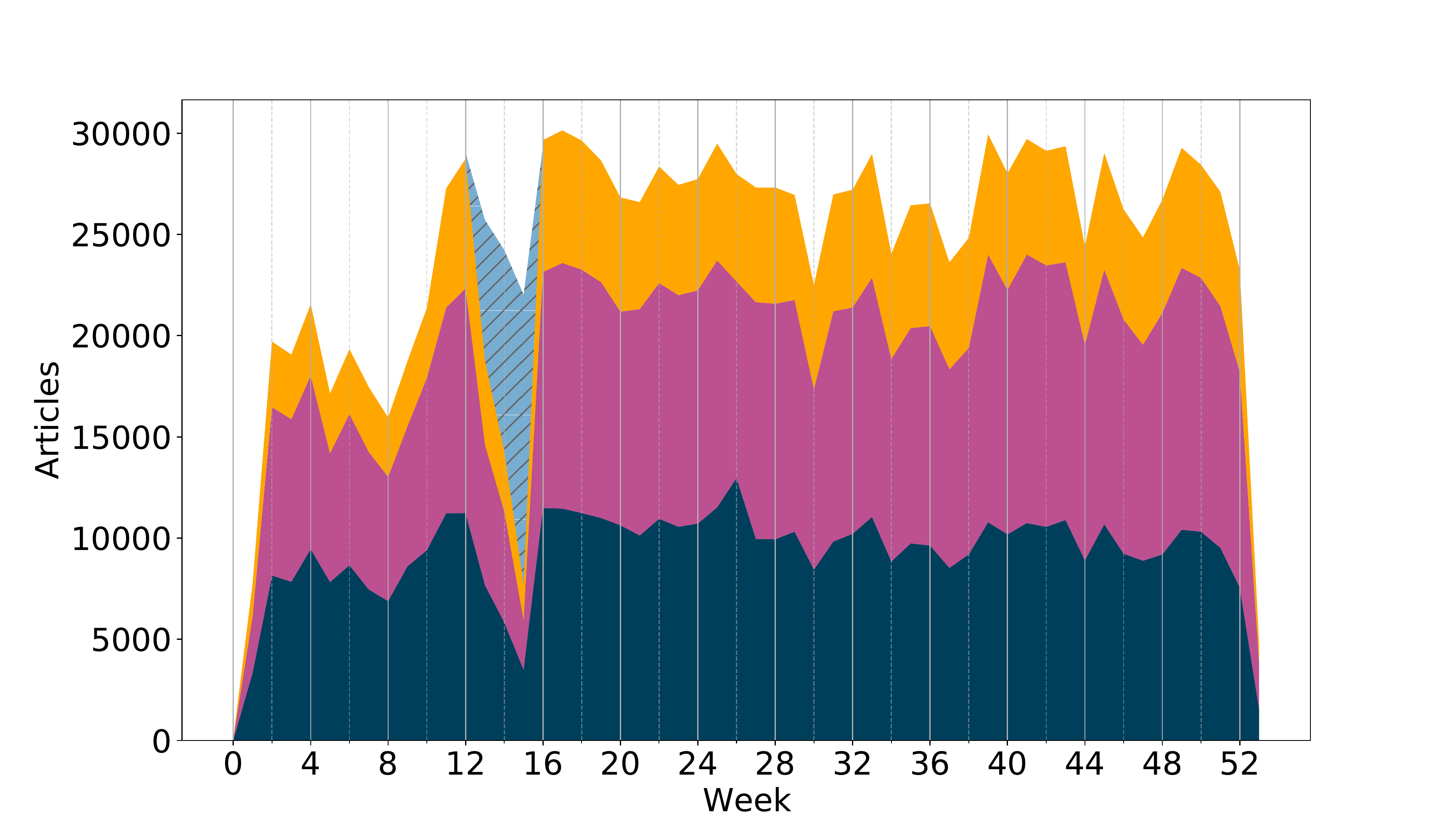}
    \caption{Number of articles per reliability class.}
    \end{subfigure} ~
    \begin{subfigure}{0.48\textwidth}
    \includegraphics[width=\textwidth]{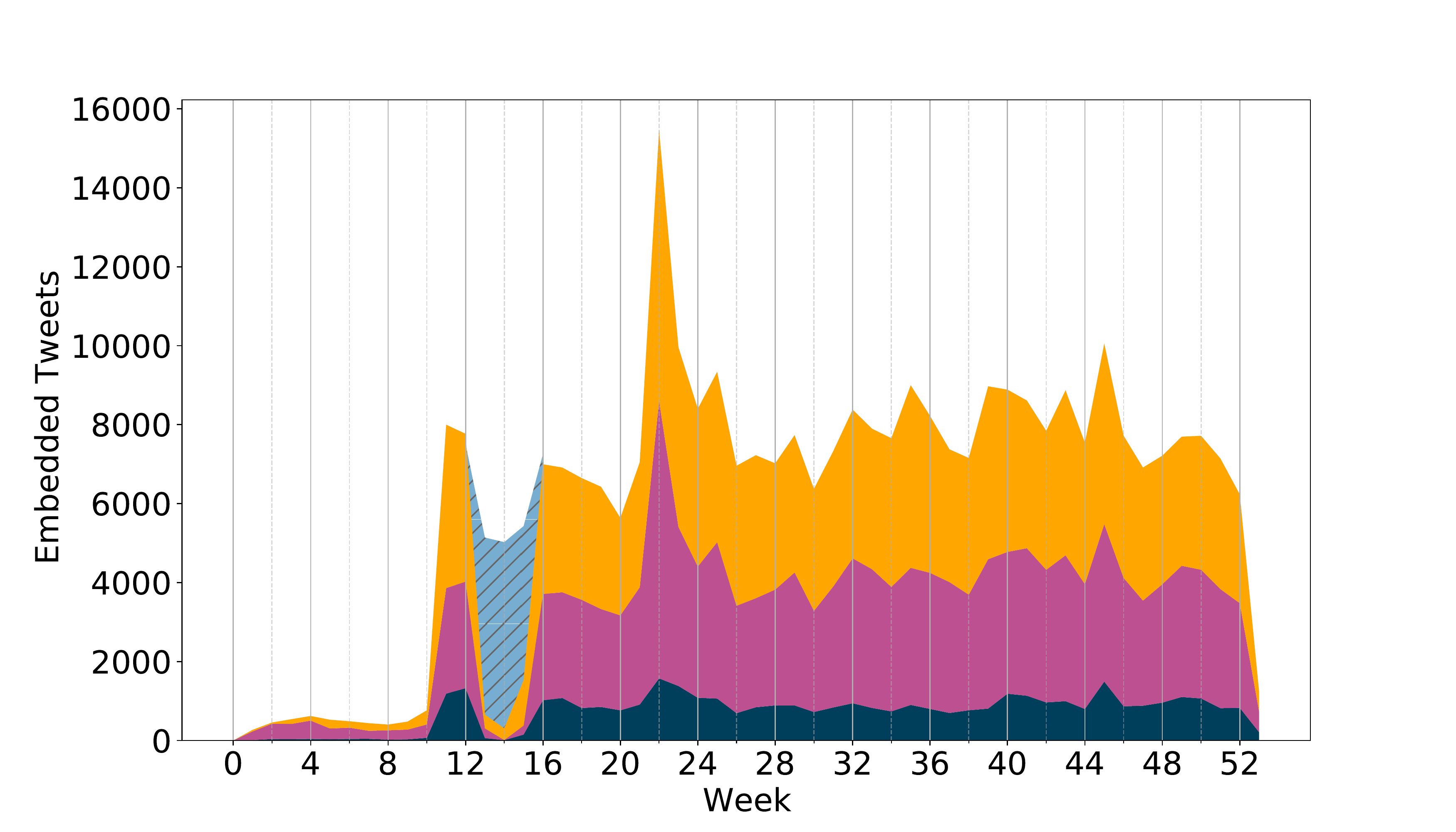}
    \caption{Number of embedded tweets per reliability class.}
    \end{subfigure}
    
    \begin{subfigure}{0.48\textwidth}
    \includegraphics[width=\textwidth]{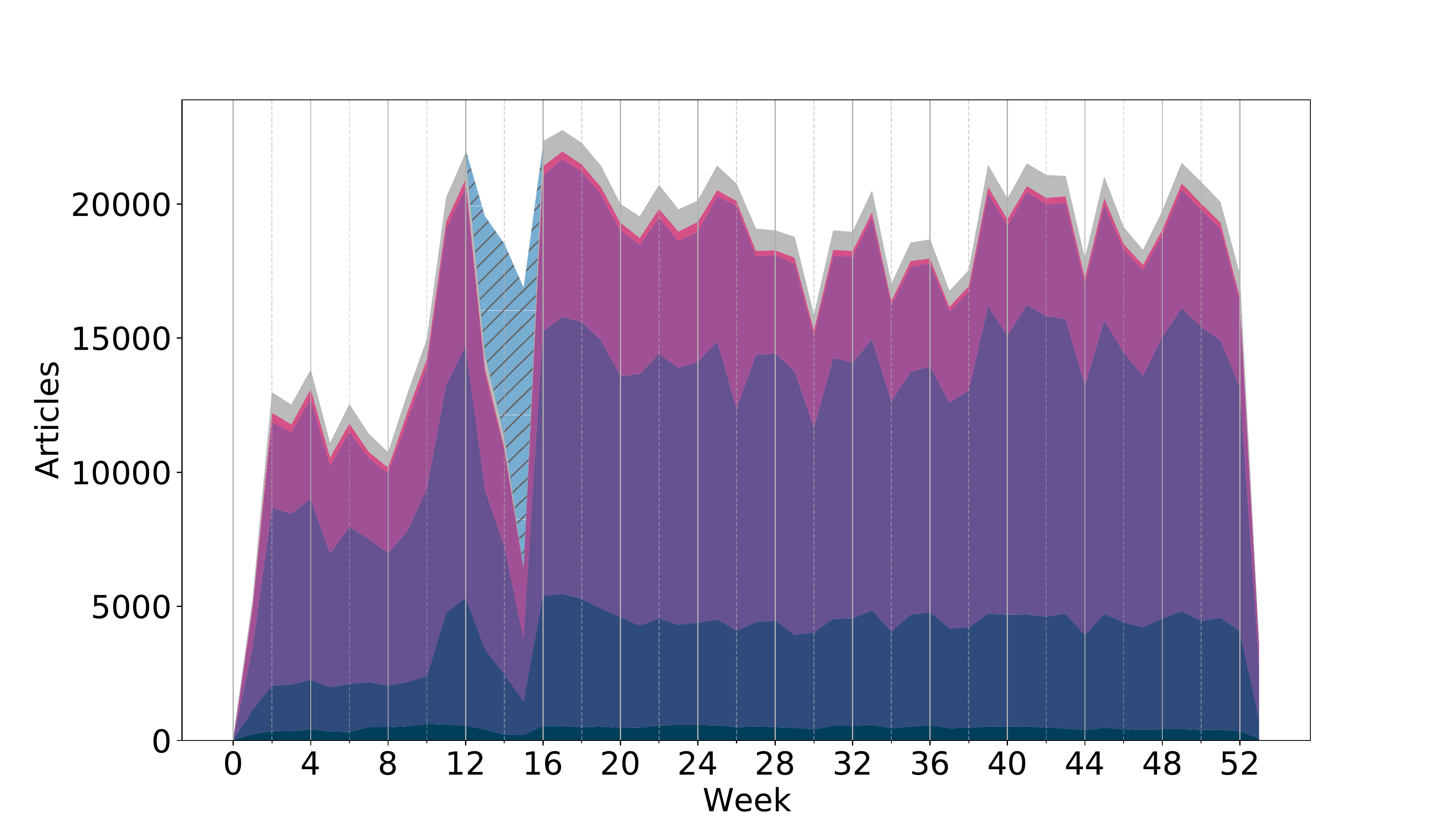}
    \caption{Number of articles per MBFC factuality score.}
    \end{subfigure}~
    \begin{subfigure}{0.48\textwidth}
    \includegraphics[width=\textwidth]{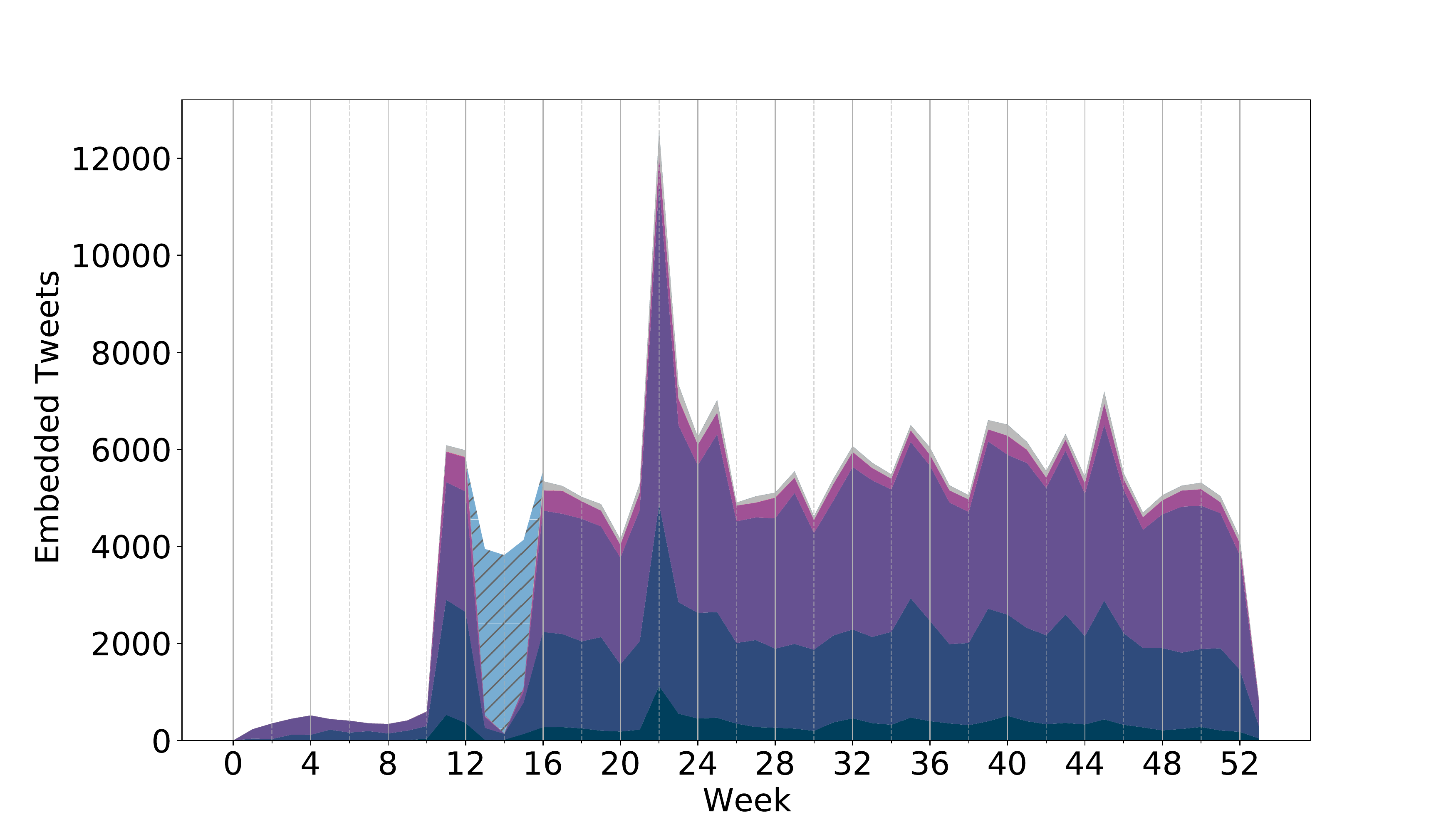}
    \caption{Number of embedded tweets per MBFC factuality score.}
    \end{subfigure}
    
    \begin{subfigure}{0.6\textwidth}
        \includegraphics[width=\textwidth]{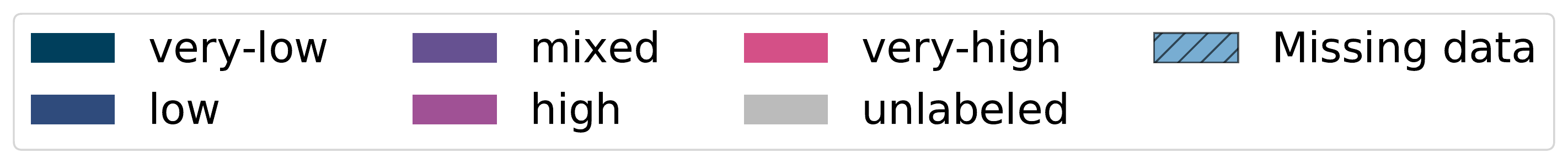}
    \end{subfigure}
    \caption{Number of articles (a, c) and embedded tweets (b, d) collected during each week of 2020.}
    \label{fig:data-weekly}
\end{figure*}

Despite having significantly improved the robustness of our collection in comparison to previous years, \texttt{NELA-GT-2020} suffered from data outage during a few weeks due to technical issues. Specifically, our collection is missing articles in weeks 13, 14, and 15, which corresponds to the the period within March $25^{th}$ through April $8^{th}$. Using linear interpolation on the data, we estimate that approximately 15,000 articles across this period were missed, this accounts for 0.8\% of the dataset. Figure \ref{fig:data-weekly} shows the collection activity in each week. The estimated missing data is shown as a gray shaded area.

\begin{figure*}
    \centering
    \begin{subfigure}{0.45\textwidth}
    \includegraphics[width=\textwidth]{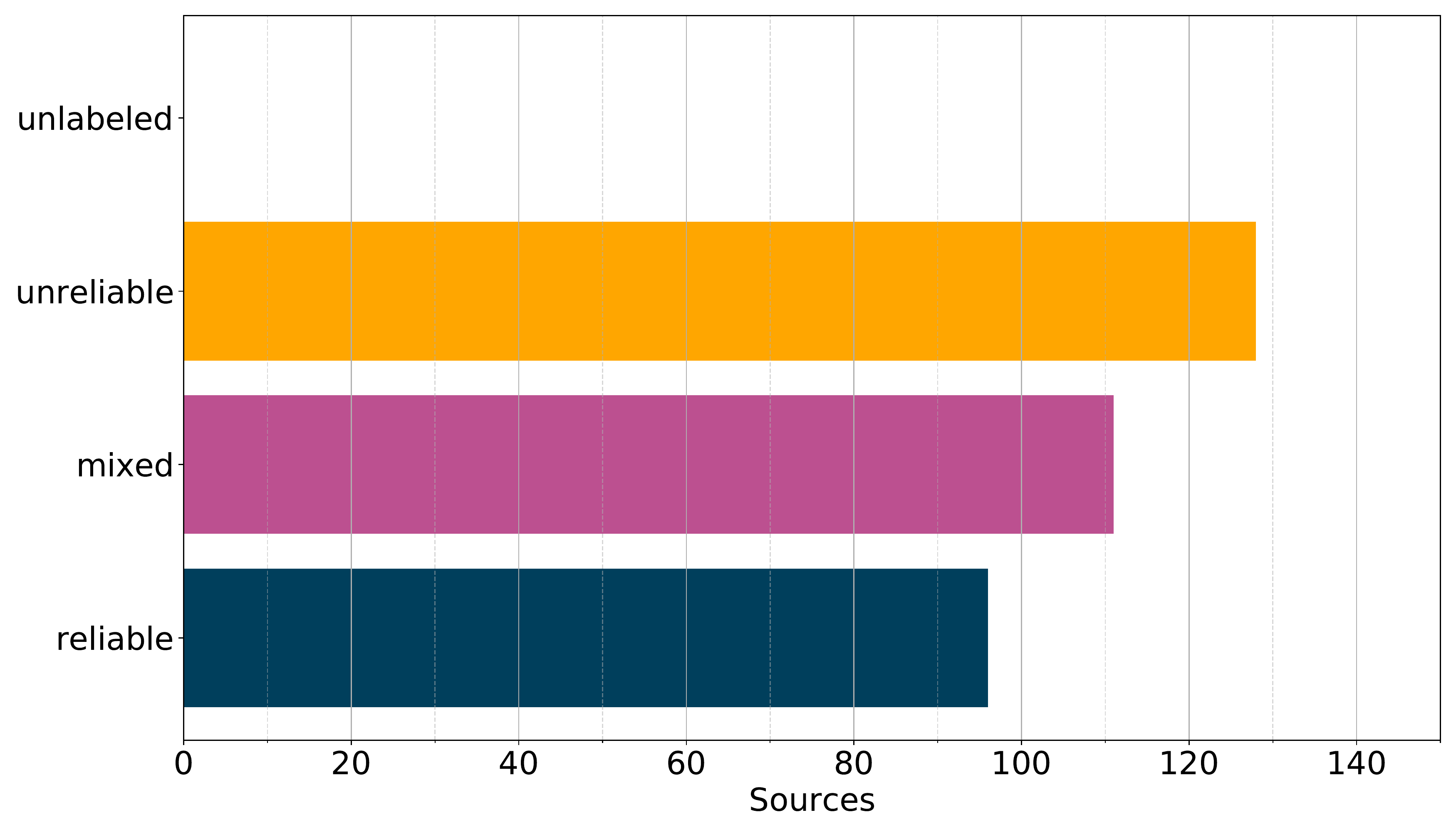}
    \caption{Number of sources per reliability class.}
    \end{subfigure} \qquad
    \begin{subfigure}{0.45\textwidth}
    \includegraphics[width=\textwidth]{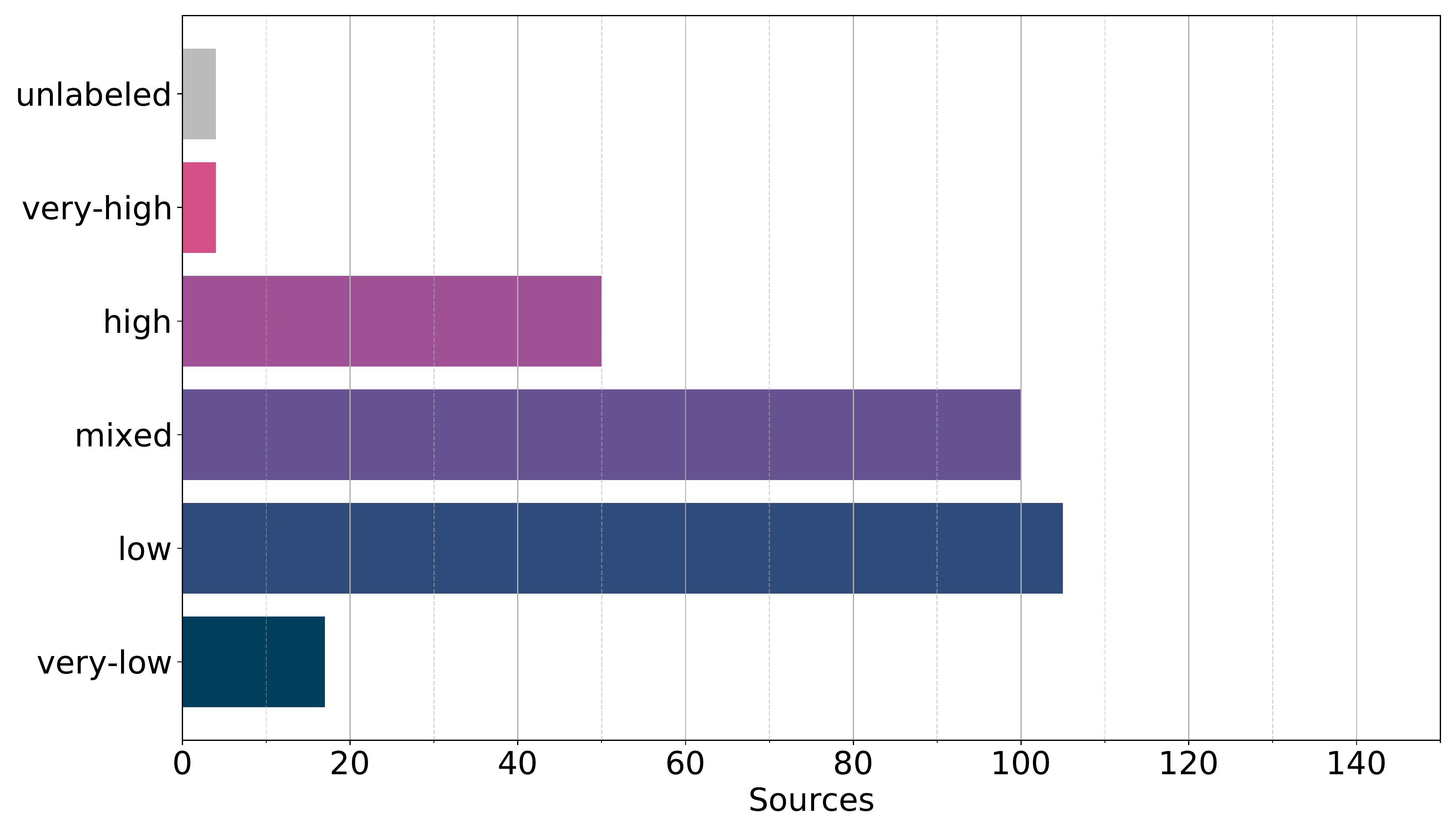}
    \caption{Number of sources per MBFC factuality score.}
    \end{subfigure}
    
    \caption{Distribution of sources per reliability class (a) and factuality (b) score.}
    \label{fig:sources-per-category}
\end{figure*}

\section{Format of Data}\label{format}
Just as in NELA-GT-2019, the dataset has been released in two formats: (1) a SQLite database, (2) a JSON dictionary per news source. Details about the structure of each of these formats is below. We provide Python code to read both data formats at: \url{https://github.com/MELALab/nela-gt}.

\subsection{SQLite Database Format}
The SQLite 3 database schema consists of twi tables: \texttt{newsdata} and \texttt{tweet}. The \texttt{newsdata} table contains, in each row, data about an article. Column \textbf{id} is set as primary key to avoid duplicated entries on the database. We normalized source names by converting them to lower case, and removing spaces, punctuation, and hyphens. For example, the source \emph{The New York Times} appears as \emph{thenewyorktimes},  Tables \ref{tab:format-newsdata} and \ref{tab:format-tweet} give information about data columns.

\subsection{JSON Format}\label{json}
We also provide the dataset in JSON format. Specifically, each source has one JSON file containing the list of all of its articles. The fields follow the same structure of the database columns (Tables \ref{tab:format-newsdata} and \ref{tab:format-tweet}).

\begin{table*}[ht!]
    \centering
    \begin{tabular}{c|c|l}
         \textbf{Column} & \textbf{Type}  & \textbf{Description}\\ \toprule
         id & text (primary key) & Article identifier. \\
         date & text & Publication date string in YYYY-MM-DD format. \\
         source & text & Name of the source from which the article was collected. \\
         title & text & Headline of the article. \\
         content & text & Body text of the article. \\
         author & text & Author of the article (if available). \\
         published & text & Publication date time string as provided by source (inconsistent formatting). \\
         published\_utc & integer & Publication time as unix time stamp. \\
         collection\_utc & integer & Collection time as unix time stamp. \\
         url & Text & URL of the article.
    \end{tabular}
    \caption{Structure of NELA-GT-2020 article data. For the database format, column \textbf{id} is the primary key of table \texttt{newsdata}.}
    \label{tab:format-newsdata}
\end{table*}

\begin{table*}[ht!]
    \centering
    \begin{tabular}{c|c|l}
         \textbf{Column} & \textbf{Type}  & \textbf{Description}\\ \toprule
         id & text (primary key) & Tweet id. \\
         article\_id & text (foreign key) & Id of the article in which the embedded tweet was observed. \\
         embedded\_tweet & text & Raw HTML of the embedded tweet.
    \end{tabular}
    \caption{Structure of NELA-GT-2020 embedded tweets. For the database format, column \textbf{id} is the primary key of table \texttt{tweet}.}
    \label{tab:format-tweet}
\end{table*}

\subsection{Ground Truth Data Format}
Just as in \texttt{NELA-GT-2019} and \texttt{NELA-GT-2018}, we include multiple types of source-level labels. In \texttt{NELA-GT-2020}, we collect source-level labels from Media Bias/Fact Check (MBFC) that contain the following dimensions of veracity:
\begin{enumerate}
    \item Media Bias Fact Check factuality score - on a scale from 0 to 5 (low to high credibility).
    \item Media Bias Fact Check Conspiracy/Pseudoscience and questionable sources - low credibility if a source belongs to these categories.
\end{enumerate}

Unlike previous versions of this dataset, we only include veracity labels from MBFC. We choose to only include these veracity labels for several reasons: 1. From our knowledge, MBFC is the most complete and most updated set of source-level veracity labels that are openly available. 2. Other external rating services, such as NewsGuard (\url{www.newsguardtech.com/}), are not freely available. We encourage researchers to corroborate the ratings from MBFC with other journalistic services if they are available. 

\section{Use Cases}
\subsection{COVID-19}
\begin{figure*}[ht]
    \centering
    \begin{subfigure}{0.4\textwidth}
    \centering
    \includegraphics[width=\textwidth]{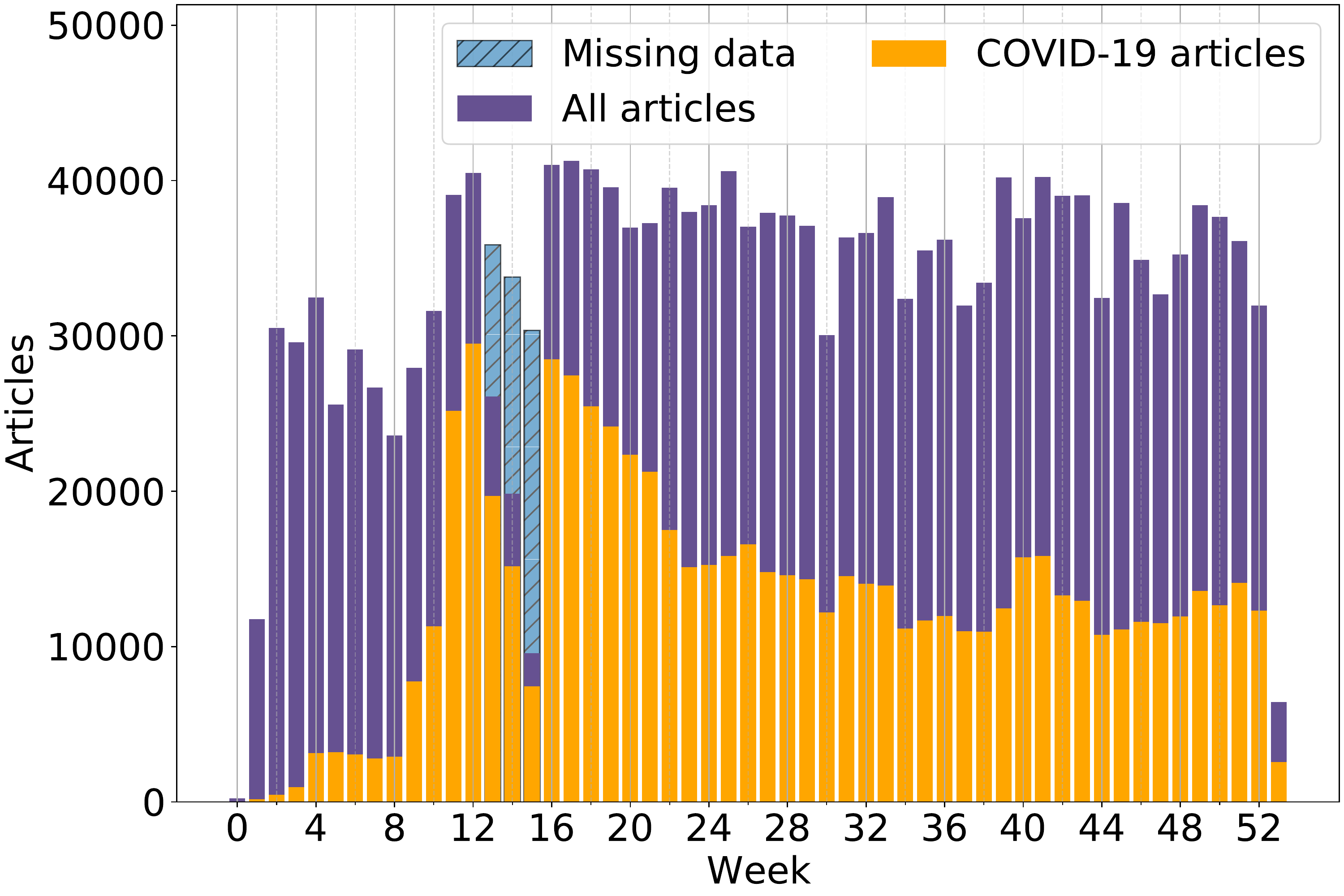}
    \caption{Number of COVID-19 articles in comparison to all articles collected in NELA-GT-2020 in each week of 2020.}
    \label{fig:all-covid}    
    \end{subfigure} \qquad
    \begin{subfigure}{0.4\textwidth}
        \centering
        \includegraphics[width=\textwidth]{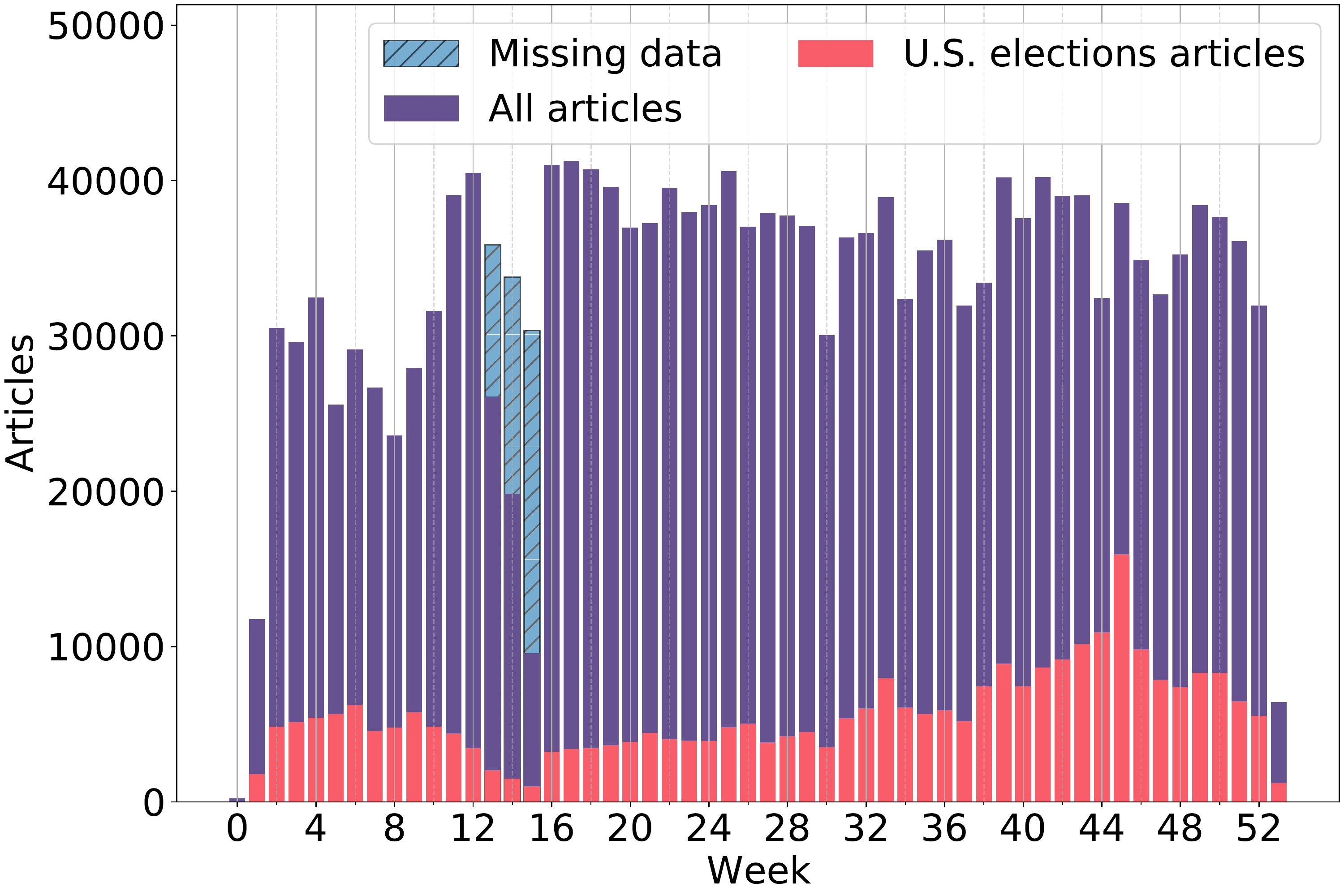}
        \caption{Number of articles related to the 2020 U.S. Presidential Election in comparison to all articles collected in NELA-GT-2020 in each week of 2020.}
        \label{fig:all-election}
    \end{subfigure}
    \caption{Number of articles related to (a) COVID-19, and number of articles related to (b) the U.S. Presidential Election as a fraction of the total number of articles in each week of 2020.}
    \label{fig:all-articles}.

\end{figure*}

One salient use case for this dataset is the study of COVID-19 media coverage. We provide a subset of the database that includes only COVID-19 related articles. This subset was generated via a simple keyword search on article title and body text. If an article had one or more keywords from our set featured in the title or body text, it was included in the COVID-19 subset. Figure \ref{fig:all-covid} shows the number of news stories related to COVID-19 compared to all articles collected in each week of 2020. 

\subsection{2020 U.S. Presidential Election}
Another salient use case for this dataset, is the study of narratives and coverage surrounding the 2020 U.S. Presidential Election. Using another simple keyword search on the article title and body text, we provide a subset of the database that only includes 2020 U.S. Election related articles. We observed that a total of 294,504 articles contained at least one election-related term, across 403 sources. Figure \ref{fig:all-election} shows the distribution of election-related articles in each week of 2020.

\subsection{Embedded Tweets}

Embedded tweets are tweets that news publishers decided to incorporate in their articles. A tweet may be the motivation of a news story, or used as evidence in the story, or may be itself the topic of the news story. We collected embedded tweets on the article page using the Goose3 library \footnote{\url{https://github.com/grangier/python-goose}}. The raw HTML code of the embedded tweet is stored in the database table \texttt{tweet}, along with the id of the article from which it was collected. 

In total, our dataset contains over 400,000 embedded tweets collected from the news articles. A single article may contain multiple embedded tweets, and a single tweet may be embedded in multiple articles. This may be used to construct a structured network of articles and tweets, which might prove, along with other signals, to be helpful in identifying signals of source reliability \cite{horne2019different,rozemberczki2019multi,gruppi2021tell,starbird2018ecosystem}.

\subsection{Long-Term Use Cases}
Just as discussed in ~\cite{gruppi2020nela}, one of our goals with the continued release of the NELA-GT datasets is to support long-term news research. When combining all of the NELA datasets, we provide over 3.5 years of news data. There are multiple research avenues that this data, both in part and as a whole, supports: 
\begin{itemize}
    \item Robust machine learning: A significant amount of work has been done in automated news veracity detection. This dataset allows for continued work in this area, particularly in robustness checks of current work. These robustness checks include examining prediction accuracy over time, over events, and over mixed veracity labels. Additionally, a wide variety of methods outside of supervised machine learning can be tested on this dataset, such as semi-supervised news veracity detection and unsupervised news veracity detection. 
    \item Exploring event-driven dynamics of and narratives in news media: Quantitative and qualitative analyses of narrative themes before, during, and after major events continues to be a useful methodology in interdisciplinary media studies. This dataset supports these works by maintaining consistent data collection across events. 
    \item Examining media manipulation: Using the veracity labels in this dataset, research can examine tactics used by purposely false news outlets. Additionally, with knowledge of media manipulation campaigns, such as those outlined in the Media Manipulation Casebook\footnote{\url{https://mediamanipulation.org/}}, researchers can examine how media manipulation is spread through malicious news outlets. While there has been a substantial focus on “fake news” detection methods by researchers, there continues to be room in understanding and characterizing media manipulation and disinformation campaign tactics. 
\end{itemize}

\begin{figure}[ht]
    \centering
    \includegraphics[width=0.9\textwidth]{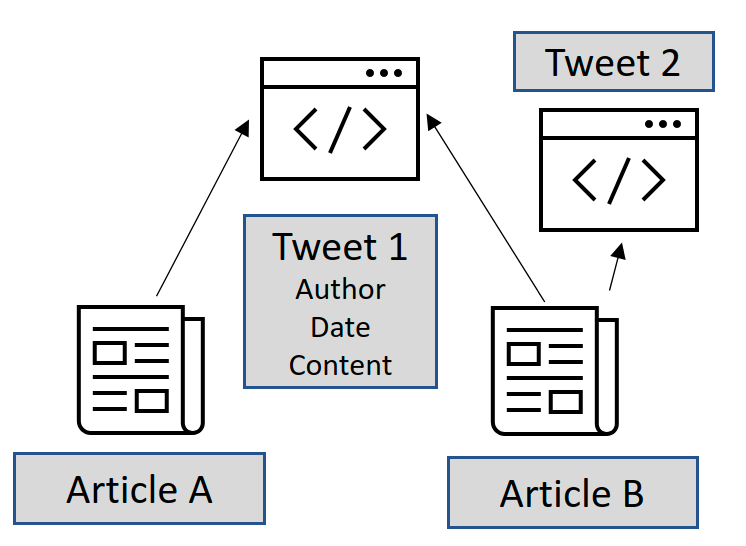}
    \caption{Relationship between news articles and embedded tweets. A single embedded tweet may be referred to by multiple articles and an article may contain multiple embedded tweets. Embedded tweets are identified by their URL.}
    \label{fig:my_label}
\end{figure}

\section{Conclusion}

With several highly controversial events, such as the many rumors about the COVID-19 pandemic and the U.S. presidential elections, 2020 was a year with plenty of opportunities for better understanding misinformation and for the development of misinformation fighting tools.
We presented \texttt{NELA-GT-2020}, a large scale dataset of news articles in English with source-level reliability labels. 
This dataset features several improvements on its predecessors \texttt{NELA-GT-2019} and \texttt{NELA-GT-2018}. First, it utilizes a more robust scraper, which is less susceptible to failures and sporadic data outages. 
The collection activity shows a much more smooth trajectory, as seen in Figure \ref{fig:data-weekly}, with no dips apart from the reported data outage in late March/early April. 
Furthermore, \texttt{NELA-GT-2020} introduces a novel feature to the collection: embedded tweets. 
We believe that this additional piece of information may be useful in modeling relationships between news articles and/or sources, as well as better understanding the relationship between news media and social media. The dataset is available in SQLite and JSON formats, minimal working examples can be found in our code repository.

\bibliographystyle{aaai}
\bibliography{main}
\end{document}